\documentstyle[epsf]{mn}
\def\epsfsize#1#2{\hsize}

\def\sgr{Sgr dSph }

\begin{document}

\title[The age of Terzan 8 ]
{The Globular Cluster System of the Sagittarius Dwarf Spheroidal
Galaxy: The age of Terzan 8 \thanks{Based on data taken at the New 
Technology Telescope - ESO, La Silla.}}

\author[P. Montegriffo et al.]
       {P. Montegriffo,$^1$ 
	M. Bellazzini,$^1$
       F.R. Ferraro,$^1$ 
D. Martins,$^2$\thanks{Visiting Astronomer, Cerro Tololo Inter-American
       Observatory. CTIO is operated by AURA, Inc., under contract to the
       National Science Foundation.} 
       A. Sarajedini$^{3,4}$\thanks{Hubble Fellow} 
\newauthor
and F. Fusi Pecci$^5$\thanks{on leave from Osservatorio Astronomico
       di Bologna}\\
        $^1$Osservatorio Astronomico di Bologna, Via Zamboni 33, 40125 
Bologna, ITALY\\
        $^2$University of Alaska, Dept. of Physics and Astronomy,
        Anchorage, Alaska, USA\\
        $^3$National Optical Astronomy Observatories, Kitt Peak 
National Observatory, Tucson, Arizona, USA \\
        $^4$San Francisco State University, Department of Physics and Astronomy,
1600 Holloway Avenue, San Francisco, California  94132\\
        $^5$Stazione Astronomica di Cagliari, Cagliari, ITALY}

\date{Accepted 
      Received ;
      in original form }
\pubyear{1997}

\maketitle

\begin{abstract}

We present deep V,I CCD photometry of the globular cluster Terzan 8, recently
found to be a member of the globular cluster system of the Sagittarius
dwarf spheroidal galaxy. We accurately estimate the metallicity of Ter 8 and
provide the first direct determination of the color excess toward this cluster.
Our robust age estimate confirms that this
cluster is indeed coeval with typical galactic globulars of comparable
metal content, and thus it is probably significantly older than at least 
two other Sagittarius clusters, Terzan 7 and Arp 2. 
The implications of this result
on the star formation history of the Sagittarius galaxy are briefly 
discussed.
\end{abstract}


\section{Introduction}

The discovery of the Sagittarius dwarf spheroidal galaxy (Sgr dSph), by
Ibata, Gilmore \& Irwin (1994, hereafter IGI-I), 
is certainly one of the most exciting
findings of the decade in the field of Galactic astronomy. 
The \sgr is the most
prominent member of the family of dwarf spheroidals orbiting the Milky Way (MW)
both in luminosity ($\sim 10^7 L_{\odot}$) and extension on the sky
(at least $22 \times 8$ degrees); in addition, it is by far the nearest 
galaxy to us (see
Ibata et al. 1997, and references therein). Because of its position
within the Galaxy and its morphology, it is commonly believed
that the \sgr is being
subjected to serious disruption by the tidal field of the MW and its destiny
will be, sooner
or later, to dissolve into the main body of the Galaxy. Furthermore, four
globular clusters, previously thought to be ordinary Galactic globulars
(i.e. NGC 6715 = M54, Arp 2, Terzan 7, Terzan 8),
are now believed to belong to the \sgr on the basis of their phase 
space coincidence
with it (Da Costa \& Armandroff 1995, IGI-I, Ibata et al. 1997), and 
at least two of them
(Arp 2, Terzan 7) are also members of a small subset of 
halo globulars whose age is significantly ounger than the bulk of 
galactic globular clusters (see Fusi Pecci et al. 1995, hereafter FBCF, 
and references therein). 

So, we are apparently witnessing the practical realization of the Searle and
Zinn (SZ) paradigm for the formation of the galactic halo (Searle \& Zinn 1978, 
Lee 1993, Zinn 1996 and references therein): a small galaxy
which has experienced a star formation/chemical enrichment history mostly
independent of the MW, and is participating in the build up of
the galactic halo, adding its own {\em young} globulars
to the MW globular cluster system. 

The relevance of this kind of mechanism to the formation of the bulk of the
halo is still under debate, and several recent results have cast some doubt
on the basic assumptions of the SZ model of galaxy formation 
(namely the very existence of an age gradient in the halo GC system, see 
FBCF, Fusi Pecci \& Bellazzini 1997 and Harris et al. 1997 for deeper insights
and references). 
Nevertheless, the above arguments make the
Sagittarius galaxy and its globular cluster system particularly 
worthy of interest. In fact, its
star formation/chemical enrichment history and its connection to the Galaxy
can help to shed light on many  mysteries 
concerning the evolution of dwarf spheroidal
systems (see Gallagher \& Wise 1994 for a review about dSph's) 
and the formation of the Galactic Halo.

For these reasons we started a long-term program
devoted to the determination of ages and chemical characteristics of
the Sagittarius system. Some important steps in this respect
have already been accomplished by our group in determining the age of Arp 2 
and Terzan 7 (Buonanno et al. 1994, Buonanno et al. 1995a, 
Buonanno et al. 1995b), even before
the very existence of \sgr was recognized. 
Moreover, photometric studies of 
the main member of the \sgr GC system  (M54=NGC6715) have recently
been published by Sarajedini \& Layden (1995), Marconi et al. (1997),
 Layden \& Sarajedini (1997) and 
 Layden \& Sarajedini (1998).

Here we present the Color Magnitude Diagram (CMD) of Terzan 8
from deep (V,I) photometry obtained with the ESO New Technology Telescope
and the CTIO 0.9 m telescope,
in order to provide a relative age estimate which is much more accurate 
than those available so far. 
The only previous study of this cluster was performed by
Ortolani \& Gratton (1990, hereafter OG90); the photometry presented here
reaches a fainter limiting magnitude and provides a much better 
definition of the principal sequences in the CMD (see sec. 4 and
compare with fig. 10b of OG90). 

The plan of the paper is as follows: in section 2 and 3,
we describe the observations and data-reduction processes, respectively;
in section 4, we present the CMD of Terzan 8 and measure its metallicity
and reddening; finally, in section 5, we derive the age of the cluster and
we discuss our results.

\section{Observations}

The data presented in this paper has been secured during two different
observational runs.

\begin{figure*}
 \vspace{20pt}
\epsffile{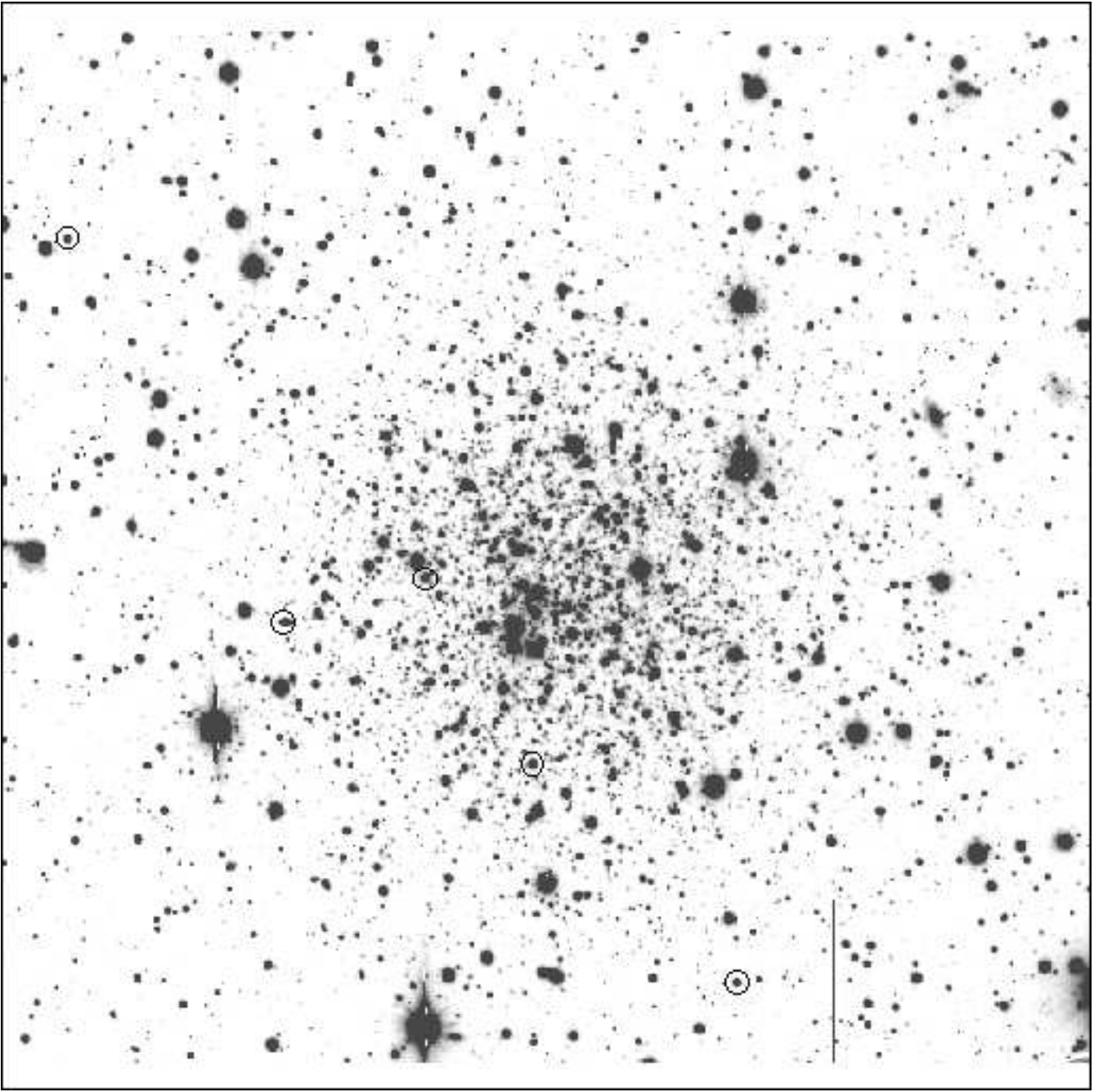}
 \caption{A map of the NTT field of view (V exposure, $t=2400 s$).
North is up, East to the left.
Candidate variables are identified by circles (see Section 3.4)}
\end{figure*}

\begin{table}
 \centering
 \begin{minipage}{140mm}
  \caption{Observations report - ESO-NTT 3.5m}
  \begin{tabular}{@{}lccc@{}}
Date          & Filter  & Exp.  & Seeing \\
              &        &  time &  (FWHM) \\
&&&\\
June 11, 1996 &  V (ESO 606) &   60 s     & $1.24^{''}$\\
June 11, 1996 &  I (ESO 610) &   60 s     & $1.05^{''}$\\
June 11, 1996 &  V (ESO 606) & 1800 s     & $1.16^{''}$\\
June 11, 1996 &  I (ESO 610) & 1800 s     & $1.05^{''}$\\
June 11, 1996 &  V (ESO 606) & 2400 s     & $1.31^{''}$\\
June 11, 1996 &  I (ESO 610) & 2400 s     & $1.27^{''}$\\
June 12, 1996 &  I (ESO 610) & 1200 s     & $1.34^{''}$\\
\end{tabular}
\end{minipage}
\end{table}
\begin{table}
 \centering
 \begin{minipage}{140mm}
  \caption{Observations report - CTIO 0.9m}
  \begin{tabular}{@{}lccc@{}}
Date          & Filter  & Exp.  & Seeing \\
              &        &  time &  (FWHM) \\
&&&\\
May 30, 1996 &  I (CTIO I31)   &  350 s     & $1.31^{''}$   \\
May 30, 1996 &  V (CTIO Vtek2) &  200 s     & $1.31^{''}$   \\
May 30, 1996 &  B (CTIO Btek2) &  420 s     & $1.43^{''}$   \\
May 30, 1996 &  I (CTIO I31)   &  350 s     & $1.18^{''}$   \\
May 30, 1996 &  V (CTIO Vtek2) &  200 s     & $1.35^{''}$   \\
May 30, 1996 &  B (CTIO Btek2) &  420 s     & $1.42^{''}$   \\
May 30, 1996 &  I (CTIO I31)   &  350 s     & $1.24^{''}$   \\
May 30, 1996 &  V (CTIO Vtek2) &  200 s     & $1.33^{''}$   \\
May 30, 1996 &  B (CTIO Btek2) &  420 s     & $1.43^{''}$   \\
\end{tabular}
\end{minipage}
\end{table}

The main set of observations were carried out 
at  ESO-La Silla (Chile) in June 1996 with the 3.5m NTT. 
In order to minimize the contamination by galactic
field stars, which is significant at the position of Terzan 8 
($l=5.76; b=-24.56$) we planned to take advantage of the high spatial 
resolution of the SUSI camera (Zjilstra et al. 1996) to probe the inner 
regions of the cluster. 
Unfortunately, the weather conditions were unsatisfactory
(two out of three nights allocated to this project
were mostly cloudy) and the 
modest seeing conditions (ranging from $1''-2''$) 
lead us to prefer the wider field of
the EMMI camera with the RILD setup (Zjilstra et al. 1996). 
This instrument provides an optically corrected
and unvignetted field of $9.15\times 8.6 ~arcmin$. The detector was a Tek
2048 CCD ($2048\times 2048$ pxs, pixel size $ 24 \mu m$) with a read-out
noise of $3.5 e^-$ rms, a dark current of $3 e^- /hr$ and a gain factor
of 1.34 $e^-/ADU$. The image scale is $0.27 arcsec/px$. 

The EMMI wide field covers almost the  whole cluster, so only one field, 
roughly centered on the cluster, was observed during the run;
its field of view is displayed in Figure 1.
 
Due to unstable weather conditions, only
frames obtained under acceptable seeing conditions
($FWHM<1.4''$) have been used in this study.
Table 1 shows the observing log for each selected frame.
Due to non-photometric conditions no standard stars were observed 
during this run (NTT).

A second set of B,V and I frames has been obtained at the 0.9 m telescope
at CTIO during the night of May 30, 1996. All of the frames were taken 
in optimal 
photometric conditions, and many standard star fields have also been 
secured. The log of observations is listed in table 2.  The detector at 
CTIO was also a Tek 2048 CCD with pixel size $ 24 \mu m$; it was operating 
at a gain setting of 3.3 $e^-/ADU$, a read-out noise of $4.4 e^-$ rms, and
a dark current of $2 e^- /hr$.  For this telescope/detector combination, the
linear scale is $0.396 ~arcsec/pix$, so the field covered was about
$13.5\times 13.5 ~arcmin$.  The CTIO CCD was controlled by the CTIO ARCON 
system, and used the standard quad amplifier mode for readout of images.

\section{Reductions}

\subsection{CTIO data}

\begin{figure}
\epsffile{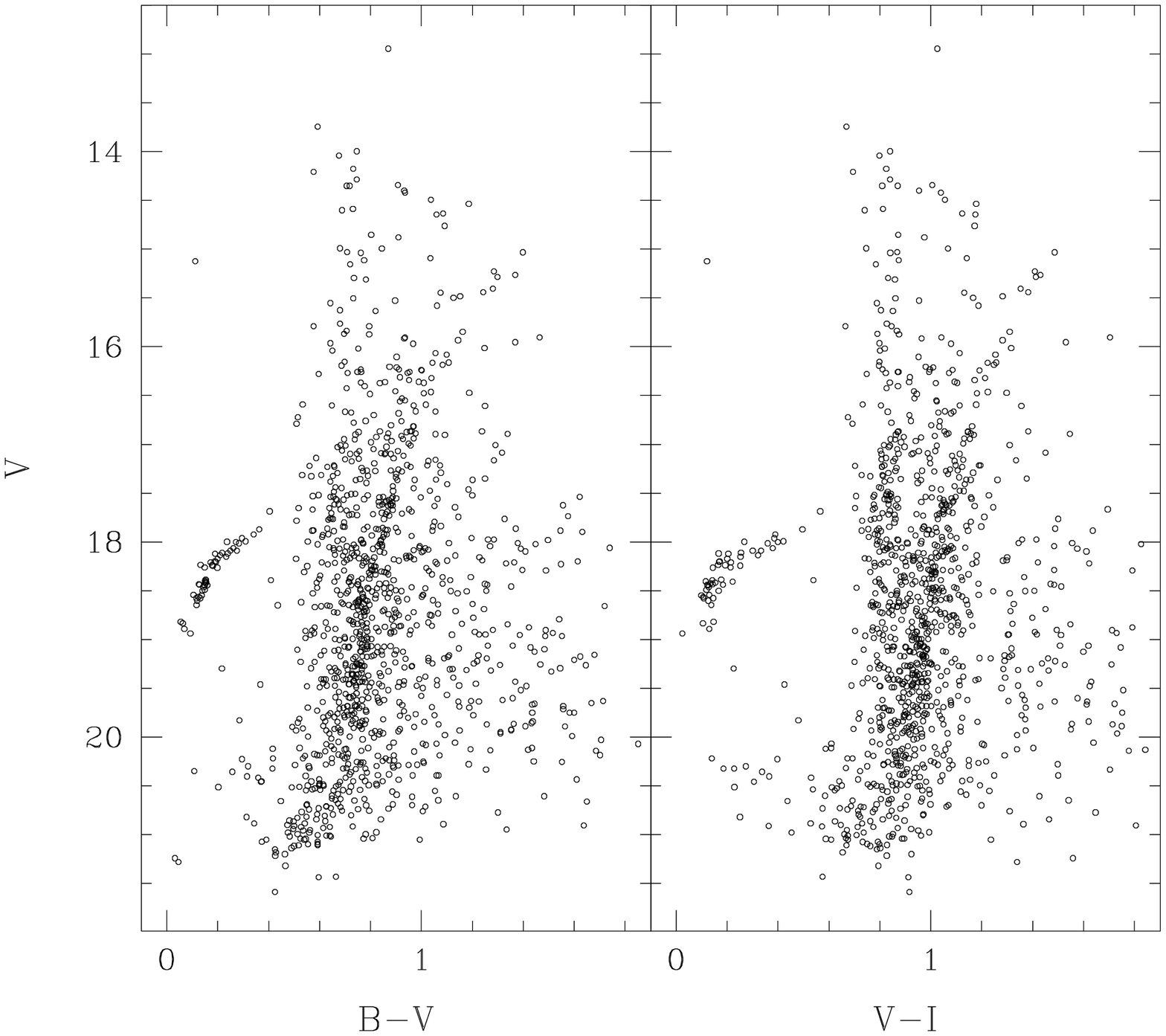}
 \caption{CMDs for the CTIO data set: (a) in (V,B-V) and (b) (V,V-I), 
respectively.}
\end{figure}

All photometric reductions of the CTIO data were done using IRAF 
\footnote{IRAF is distributed by the National Optical Astronomy
Observatories which are operated by the Association of Universities
for Research in Astronomy, Inc. under cooperative agreement with
the National Science Foundation.} on
a dual 200 Mhz Pentium Pro system running Solaris 2.5 for x86 processors
at the University of Alaska in Anchorage.  Preprocessing of the CTIO
material was carried out at Kitt Peak National Observatory using 
standard IRAF/QUADPROC procedures.
The CTIO data were used to provide a photometric calibration for the NTT
observations.  This was done by first establishing the photometric
transformations to the V, B-V, V-I system using standard star observations
of selected fields in the list by Landolt (1993).  A total of 38 different 
stars were observed from several fields, covering a wide range of colours.
Extinction was well determined since the standard star
observations covered a range from 1.1 to 2.1 airmass.

Reduction of the standard star data was done using standard processing within
the IRAF/DAOPHOT/PHOTCAL routines.  The B, V, and I equations were
fit separately in this processing, yielding the following transformation 
equations:

$$b = B + 3.203 + 0.228X + 0.084(B-V)$$ 
$$v = V + 2.967 + 0.123X - 0.027(B-V)$$ 
$$i = I + 3.914 + 0.035X - 0.025(V-I)$$ 

\noindent where X is the airmass;, b, v, and i are the instrumental
observations, and B, V, and I are the Landolt (1993) standard values. The
root mean square deviation of the standard values from these fits are 0.018 mag,
0.012 mag, and 0.024 mag, respectively.

The cluster frames were processed individually rather than averaging
exposures in each filter.  A spatially variable point spread function
(PSF) was determined for each frame using a single set of 150 stars  
and each frame was processed with DAOPHOT/ALLSTAR.  The aperture correction
was accomplished using a single set of 14 stars observed in all frames
with ALLSTAR as well as with PHOT using the same configuration employed
with the photometric standard stars.

However, checking the variation of the aperture correction 
as a function of X, Y , and radial position (R) relative to the  
cluster center revealed a quadratic dependence of the aperture
correction on R.
This variation appears to be due to the spatially variable PSF, which
is itself caused by curvature in the surface of the CCD, optical 
problems in imaging, etc. The effect is small in this case, amounting
to corrections on the order of 0.0 to 0.03 mag over the field, except
for V which displayed a slightly larger variation.

A quadratic equation was fitted to the spatially variable aperture
correction and applied to the ALLSTAR output magnitudes. The resulting
instrumental photometry was transformed to the standard system and
combined in the following manner.  The nine frames were grouped in
three sets of B, V, I images, so the final result was three sets of
V, B-V, V-I data.  These three data sets were averaged, with
rejection of stars which were not found in all three sets within 
one pixel of each other, and with any star rejected if the V magnitude
varied by more than 0.05 magnitude.
This final, averaged result was then used to calibrate the NTT data
as described below.

Figure 2(a,b) shows the (V, B-V) and (V, V-I) diagrams obtained from the
CTIO observations. Because of the wide field, the contamination by galactic
field stars is quite high in these diagrams. Nevertheless the RGB and
the blue HB are clearly identifiable in both diagrams.

\subsection{NTT data}

All of the reductions were carried out using the DAOPHOT package 
(Stetson 1987) mounted on a Digital-Alpha station of the Bologna Observatory.
In order to determine the PSF and to find and fit stars in each frame,
we adopted the standard procedure as described by Stetson (1987).

The program stars were detected automatically adopting a threshold-criterion
($\sim 4 \sigma$) above the local background level.
The searching procedure was performed independently on each frame
listed in Table 1. In particular we used the shortest exposures
($t=60 s$) to properly measure the ($\sim 40$) brightest stars 
near the RGB tip, which were heavily saturated on the deeper exposures.

The {\it instrumental} magnitudes in each colour obtained in each frame were
referred to the best quality frame assumed to be the {\it reference},
and then averaged properly weighting the photometric quality of each frame.
Since our main goal was to derive a good estimate of the age and metallicity
of the cluster using the principal sequences in the CMD,
we optimized the process of data reduction in order to obtain a final
CMD which is as {\it clean } as possible.
For this reason, we used the image quality diagnostics in DAOPHOT
to discard stars with poor measurements; no attempt was made to 
estimate the degree of completeness of the sample.
At the end of the reduction procedure a set of {\it instrumental}
magnitudes, colours and positions was obtained for a sample of 3115 stars.

Since no photometric calibration was obtained during the NTT run, we used
the well calibrated CTIO set to link the 
NTT {\it instrumental} magnitudes 
to the standard Johnson system.

\begin{figure}
\epsffile{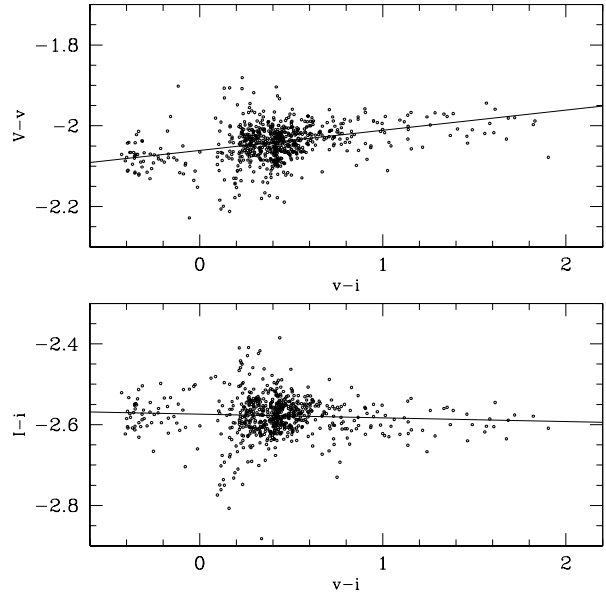}
\caption{Magnitude difference as a function of the {\it intrumental}
colour $(v-i)$ for the 684 stars
in common between the NTT and CTIO data sets. 
The lines are the least squares fit to the 
data and represents the 
equations relating the two sets of observations.}
\end{figure}

The cross correlation between the NTT  and the CTIO  catalogues
identified 684 stars in common which were used to calibrate the NTT data.
The equations relating the NTT {\it instrumental} system
to the CTIO calibrated data set are:
$$V=v-2.061+0.050 (v-i)$$
and
$$ I=i-2.574-0.009 (v-i)$$
as derived from the least squares fit (with a recursive 2 $\sigma $~ 
rejection) of magnitude residuals as a function of the NTT {\it instrumental} 
$(v-i)$ colour.
The calibration curves are plotted in Figure 3.
As can be seen, the stars in common cover a sufficiently wide range in colour
which should prevent any residual uncorrected colour trend.

The list of the final calibrated $V,I$ magnitudes
and positions for 3115 star identified in the field covered by our 
observations and which satisfy the above selection criteria, is available
upon request to the first author\footnote{Send e-mail request to 
montegriffo.@astbo3.bo.astro.it.}.
Figure 4 shows the CMD obtained from the whole NTT sample
of measured stars.

\subsection{Photometric errors}

A first-order estimate of the internal accuracy of the photometry
can be obtained from the rms frame-to-frame scatter of the instrumental 
magnitudes, once they have been referred to the standard
{\it  reference frame} as
described above.

The rms values has been computed for each individial star
in the NTT data set according to the formula:

$$ \sigma_{s}=\biggl [{{{\Sigma_{i=1}^{N_s} w_{si} (v_{si} - <v_{si}>)^2}
\over{(N_s-1)\Sigma_{i=1}^{N_s} w_{si}}}\biggr ] ^{1\over{2}}} $$

\noindent
where $w_i={\sigma_i}^{-2}$ and ${\sigma_i}$ are the DAOPHOT photometric
internal errors (Stetson 1987), $N_s$ is the number of independent magnitude
estimates for the $s-th$ star, $v_{si}$ is the instrumental magnitude, and
$<v_s>$ is the weighted mean instrumental magnitude of the stars.
For the $\sim 40$ brightest stars, the only ones that were measured only
on a single couple of frames (i.e. the {\em short} V and I exposures),
we assumed $\sigma_s = \sigma_i$.

The rms values obtained for the V and I filters are plotted in Figure 5a,b
as a function of the mean calibrated magnitude adopted for each star.
The mean errors for $13<V<20$ are less than 0.01 mag, and they are however
less than 0.1 mag even at the faint limit.

\subsection{Variable stars}

The {\it rms} frame-to-frame scatter can be used to obtain 
 indications on variability of the objects identified by
our reduction procedure.
In particular, the detection of any RR Lyrae variable can be important to
clearly define the exact HB morphology 
of Ter 8 ({\it i.e} to establish whether all HB
stars are bluer than the instability strip).
From the analysis of Figure 5 it is evident that there are some stars which
show rms values significantly
higher than the mean value expected at the HB level 
($V \sim 18$): these stars might be  variables.
On the basis of the  {\it rms} scatter we identified $\sim 10$ candidates. 
Among them,
five stars (namely \# 117, 688, 1350, 1582, 2784)
show large differences in magnitude also with respect to the CTIO
measures, confirming they might be {\it true} variables.
Their position (in pixels) and magnitude are listed in Table 3 (they are also
 identified in the map shown in Figure 1).

Our photometry does not have a sufficient time coverage to properly
define a light curve,
so the evidences presented here should be considered
as preliminary indications of variability. A specific
follow-up study is need to firmly establish their nature.

\begin{figure}
\epsffile{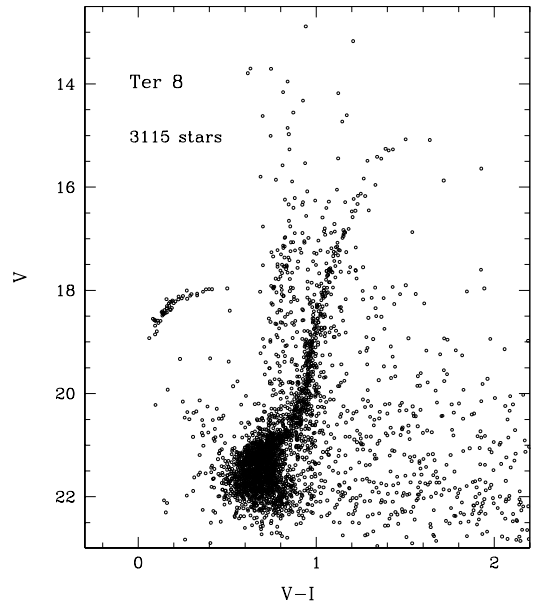}
 \caption{Color Magnitude Diagrams from the stars
observed in Terzan 8 in the NTT frames.}
\end{figure}

\begin{table}
 \centering
 \begin{minipage}{140mm}
  \caption{Magnitudes and positions for candidate variables.}
  \begin{tabular}{@{}ccccc@{}}
  Name  &     V   &   I   &    X    &   Y    \\ 
   117  &  18.03  &  17.34 & 1381.9 & 152.6\\
   688  &  17.55  &  16.29 &  998.4 & 563.0\\
   1350  &  17.96  &  17.46 &  531.8 & 830.0\\
   1582  &  17.27  &  16.35 &  796.3 & 913.1\\
   2784  &  17.61  &  16.39 &  122.1 & 1551.1\\
\end{tabular}
\end{minipage}
\end{table}

\section{ The Color-Magnitude Diagrams}

The inspection of the CMDs presented in Figure 2 and 4 shows that
the field contamination by the Galaxy ($0.5<(V-I)<1$) is particularly 
severe at the latitude of Terzan 8. Despite this, the principal
sequences of the cluster population in the CMD are quite well defined.
The overall morphology of the CMD can be summarized as follows, as it appears
from Figure 4.

\begin{itemize}

\item{} The Red Giant Branch is well defined, though sparsely populated 
in the very upper portion. The brightest measured giant is
located at $V\sim15$ and $V-I\sim1.5$.

\item{} The Horizontal Branch (HB) is populated mainly on the blue side of 
the instability strip as expected for an old, metal poor population
(Da Costa \& Armandroff 1995, OG90) and the blue tail extends down
to $V\sim 19$.

\item{} The subgiant branch (SGB) and the upper main sequence are defined 
and well populated down to $V\sim 22.5$ and the main sequence turnoff
(TO) can be located at $V_{TO}=21.55\pm0.10$, as measured from the derived
ridge line and from the {\em rms} dispersion of the points around it
in the TO region (see below).

\end{itemize}

In order to compute the mean ridge lines of the main branches of
the CMD plotted in Figure 4 we followed an iterative process.
First, the portions of the CMD containing the cluster branches were 
selected by eye. Then, we adopted two different methods
depending on the magnitude:

\begin{enumerate}

\item{} {\it the bright region of the CMD (i.e. for $V<20$, RGB and HB)}:
we adopted the polynomial fitting technique described by Sarajedini \& 
Norris (1994). The main branches have been iteratively fitted by 
a second order polynomial; in each iteration stars more 
distant than
$2 \sigma$ from the fit were rejected and the fitting procedure
repeated until the solution was stable, i.e. no more stars were rejected.

The resulting equation was:
$$ (V-I) = 1.043 -0.083 \times (V-18) + 0.017 \times (V-18)^2 $$

\item{} {\it the faint region of the CMD (i.e. $V>20$, SGB and upper-MS)}:
the stars were divided into 0.4-mag wide bins in V magnitude.
The median color and magnitude of each bin was then computed and
the stars more distant than $2 \sigma$ from the median point of the 
bin were rejected. The process was repeated until the difference
of the median between two successive iterations became negligible.
The {\it median points} were then interpolated with a cubic spline. 

\end{enumerate}

The resulting ridge lines obtained as described above match smoothly at
$V=20$ and 
have been tested to be very stable against different
choices of the initially selected area.
They provide an excellent representation of the branches of Terzan 8
CMD; the adoped normal point of the main branches are listed in Table 4.

\begin{figure}
\epsffile{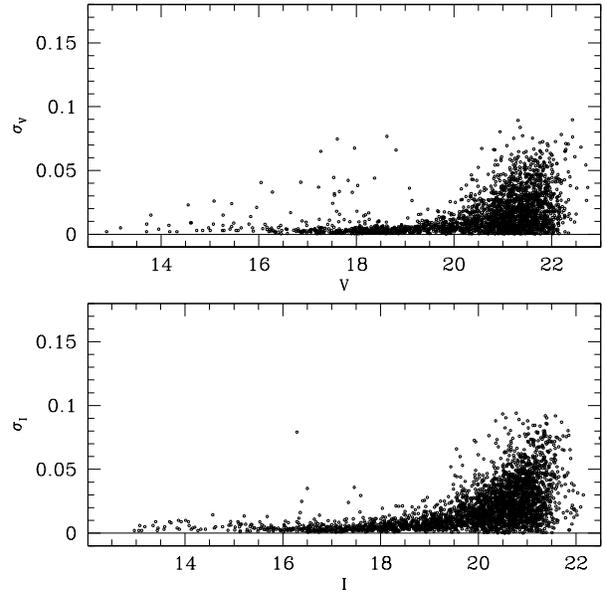}
 \caption{Rms values of the frame-to-frame scatter, in magnitude are plotted
versus the mean adopted magnitude for each measured star.}
\end{figure}

\subsection{The level of the Horizontal Branch}
 
The horizontal part of the cluster HB is scarcely populated and
falls in a region of the CMD where the field contamination is
significant. To derive the $V_{HB}$\footnote{In the following we 
define $V_{HB}$ as the mean V
magnitude of the RR Lyrae variables. 
In all cases in which a detailed study of the cluster RR Lyraes
is not available, the mean V magnitude of
the stars located in the instability strip is adopted as the estimator of
$V_{HB}$ (see Chaboyer, Demarque \& Sarajedini, 1996, hereafter CDS).}
we followed the procedure 
already adopted in previous papers (see Ferraro et al. 1992; Ferraro, 
Fusi Pecci \& Buonanno 1992; Sarajedini 1994; Cassisi \& Salaris 1997).

We considered reference clusters with {\em a)} roughly comparable metallicity,
{\em b)} a HB populated both on the blue and red side of the RR Lyrae
region and {\em c)} a precise determination of the HB
level is available from RR Lyrae analysis; then we shifted the 
Terzan 8 CMD until the
HB sequence overlaps the blue HB of the reference cluster.

Since Terzan 8 is a quite metal poor cluster ($[Fe/H]=-1.99 \pm 0.08$
Da Costa \& Armandroff, 1995), we
adopted as reference clusters two well-studied metal poor clusters, 
i.e. M68 ($[Fe/H]=-2.09$) and M15 ($[Fe/H]=-2.15$). 
A detailed study of the RR Lyrae population in M68 
has recently been published by Walker (1994) who found $V_{HB}=15.64\pm0.01$.
In figure 6 (panel a) the HB stars for Terzan 8 in the (V,V-I) plane
(filled squares) have been shifted 
($\Delta V=-2.32$ and $\Delta (V-I)=-0.1$) to match the blue HB of 
M68 (open squares): the solid line is the $V_{HB}$ level found by 
Walker (1994). This procedure yields $V_{HB}^{Ter8}=17.96$.

M15 is a metal poor cluster with a large population of RR Lyraes
(more than 100 has been listed by Sawyer-Hogg, 1973, but new candidates have 
been found by Ferraro \& Paresce, 1993 and Butler et al. 1997). 
An accurate study of the 
RR Lyrae in M15 has been published by Bingham et al. (1984) who found 
$V_{RR}^{M15}=15.83\pm 0.01$. The zero point of the Bingham et al. (1984)
calibration is consistent with the photographic study by 
Buonanno et al. (1983). For this reason, Figure 6b presents the HB stars in 
Terzan 8 in the (V,B-V) plane (from the CTIO observations)
(filled squares) shifted (by $\Delta V=-2.12$ and $\Delta (V-I)=-0.09$)
to match the blue HB tail of M15 (open squares) by Buonanno et al. 1983. 
This procedure yields $V_{HB}^{Ter8}=17.95$.
The heavy solid line is the $V_{RR}$ level found by Bingham et al. (1984).

The two determinations 
turns out to be highly consistent; they 
differ by only one hundreth of a magnitude. OG90 do not provide an explicit
extimate of $V_{HB}$ but inspection of their CMD (OG90, fig. 9) suggests
a good agreement with our measures. 
In the following, we adopt 
$V_{HB}=17.95 \pm 0.05$, the errors being a conservative 
estimate of the uncertainty of the procedure.

\begin{figure}
\epsffile{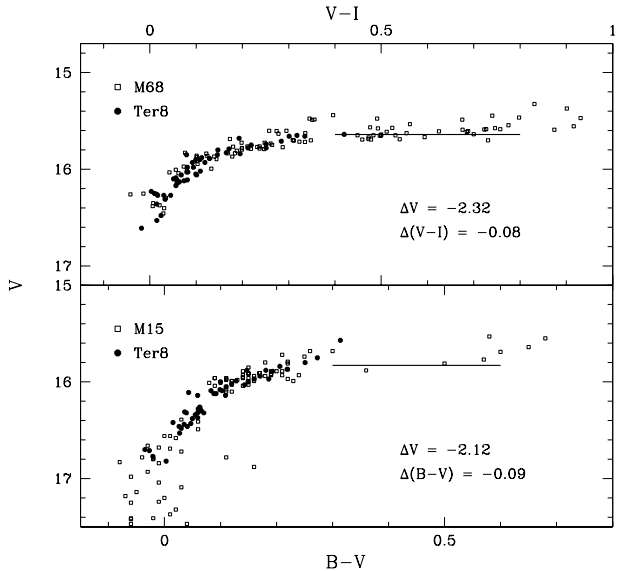}
 \caption{HB stars in Terzan 8 (small filled circles)
have been shifted to fit:
{\it (panel a)}
 the HB in M68 
by Walker (1994) (empty circles) in the (V,V-I) plane.
The heavy solid line is the RR Lyrae level in M68;
{\it (panel b)}
the HB in M15 
by Buonanno et al. (1983) (empty circles) in the (V,B-V) plane.
The heavy solid line is the RR Lyrae level in M15 (Bingham et al. 1984).}
\end{figure}

\subsection{Metallicity and Reddening}

As is well known the shape and position of the RGB can be
used to derive a {\it photometric} estimate of the cluster metallicity
and reddening. In order to get a simultaneous measure of $[Fe/H]$ and 
$E(V-I)$, we apply the SRM method developed by Sarajedini (1994)
in the (V $vs.$ V-I) plane, using the 
$(V-I)_{g}$ and the RGB ridge line listed in Table 4
(see Sarajedini 1994 for definitions and details).
The same procedure can also be applied in the (V $vs.$ B-V) plane
using the appropriate calibration (Sarajedini \& Layden 1997).

Assuming $V_{HB}=17.95\pm 0.05$ (see Section 4.1)
we find $(V-I)_g=1.050\pm0.013$ and $(B-V)_g=0.840\pm0.013$, 
respectively, in the (V,V-I) and (V,B-V) planes.
With these values and the mean ridge lines listed in Table 4 
the SRM method yields:

$[Fe/H]=-2.03\pm0.14$ and $E(B-V)=0.12\pm0.02$ 

$[Fe/H]=-1.98\pm0.13$ and $E(B-V)=0.12\pm0.03$, 

\noindent
from (V,V-I) and (V,B-V) CMD, respectively. The relation 
$E(V-I)=1.3\times E(B-V)$ by Dean et al. (1987) has been adopted
in order to convert E(V-I) into E(B-V).

The errors associated with the derived quantities include the combination
of three main uncertanties:

\begin{enumerate}

\item{} the uncertainty of $0.05 mag$ in the determination of the magnitude 
level of the HB (see the previous section);

\item{} the uncertanty in V-I of the ridge line at the HB level,
estimated to be $0.013 $mag, i.e. the local standard deviation from
the ridge line.

\item{} the uncertainty in B-V of the ridge line at the HB level,
estimated to be $0.013 mag$, i.e. the local standard deviation from
the ridge line.

As can be seen, the two semi-independent values derived for the 
metallicity and reddening are in excellent agreement. Moreover,
the estimated metallicities show remarkable agreement with previous 
determinations, 
both photometric ($[Fe/H]\sim -2.0$, OG90) and spectroscopic
($[Fe/H]=-1.99 \pm $Da Costa \& Armandroff 1995). 

The weighted means of the two values yield
$[Fe/H]=-2.00\pm0.10$ and $E(B-V)=0.12\pm0.03$, where,
besides the formal errors, the quoted errors take into account
a conservative estimate of the uncertainty due to the sensitivity of 
the method to the details of the assumed ridge line 
especially in the case of metal poor clusters with very steep RGBs.

In summary
our results provide: {\em a)} the first direct measure of the reddening
for this cluster and {\em b)} a confirmation that the metallicity 
as measured by photometric parameters 
coincides with that found with spectroscopy
of cluster giants. 

So it seems that Terzan 8 does not suffer the same
syndrome as the Sagittarius clusters (i.e. Ter 7) 
and the young clusters (Pal 12), whose spectroscopic
measures yield a higher metal content than the photometric ones
(by $\sim 0.2 - 0.4 ~dex$; see FBCF, Sarajedini \& Layden 1997, Da Costa \&
Armandroff 1995). Indications for a {\it decoupling} of spectroscopic
and photometric metallicity indexes has been found also for Arp 2 and Rup 106
(FBCF,Sarajedini \& Layden 1997)
though the differences
are not inconsistent within the measurement errors. However, the high dispersion
spectroscopy  study of Rup 106 giants by Brown, Wallerstein \& Zucker (1997)
revealed that this cluster do indeed present significant chemical peculiarities
with respect to a typical Halo population. So it seems not excluded that
the spectroscopic {\it vs.} photometric differences in the measurement of 
clusters metallicity could track true differences in the chemical mix.

\end{enumerate}

\begin{table}
 \centering
 \begin{minipage}{140mm}
  \caption{Terzan 8 :fiducial points}
  \begin{tabular}{@{}cccccccc@{}}
  &{\bf RGB} &   & &  & &  &  \\
  V-I  &     V   &   V-I   &    V    &   V-I   &    V    &   V-I   &    V \\ 
  1.485  &  14.80  &  1.088  &  17.50  &  0.941  &  20.10  &  0.675  &  21.45 \\
  1.466  &  14.90  &  1.079  &  17.60  &  0.939  &  20.15  &  0.674  &  21.50 \\
  1.447  &  15.00  &  1.069  &  17.70  &  0.936  &  20.20  &  0.674  &  21.55 \\
  1.429  &  15.10  &  1.060  &  17.80  &  0.934  &  20.25  &  0.674  &  21.60 \\
  1.411  &  15.20  &  1.051  &  17.90  &  0.932  &  20.30  &  0.675  &  21.65 \\
  1.393  &  15.30  &  1.043  &  18.00  &  0.928  &  20.35  &  0.676  &  21.70 \\
  1.375  &  15.40  &  1.035  &  18.10  &  0.924  &  20.40  &  0.679  &  21.75 \\
  1.358  &  15.50  &  1.027  &  18.20  &  0.918  &  20.45  &  0.681  &  21.80 \\
  1.341  &  15.60  &  1.019  &  18.30  &  0.909  &  20.50  &  0.685  &  21.85 \\
  1.325  &  15.70  &  1.012  &  18.40  &  0.896  &  20.55  &  0.688  &  21.90 \\
  1.309  &  15.80  &  1.005  &  18.50  &  0.880  &  20.60  &  0.692  &  21.95 \\
  1.293  &  15.90  &  0.999  &  18.60  &  0.860  &  20.65  &  0.697  &  22.00 \\
  1.278  &  16.00  &  0.993  &  18.70  &  0.836  &  20.70  &  0.702  &  22.05 \\
  1.263  &  16.10  &  0.987  &  18.80  &  0.817  &  20.75  &  0.707  &  22.10 \\
  1.248  &  16.20  &  0.981  &  18.90  &  0.798  &  20.80  &  0.712  &  22.15 \\
  1.234  &  16.30  &  0.976  &  19.00  &  0.779  &  20.85  &  0.718  &  22.20 \\
  1.220  &  16.40  &  0.972  &  19.10  &  0.761  &  20.90  &  0.723  &  22.25 \\
  1.206  &  16.50  &  0.967  &  19.20  &  0.744  &  20.95  &  0.729  &  22.30 \\
  1.193  &  16.60  &  0.963  &  19.30  &  0.730  &  21.00  &  0.735  &  22.35 \\
  1.180  &  16.70  &  0.960  &  19.40  &  0.718  &  21.05  &  0.741  &  22.40 \\
  1.167  &  16.80  &  0.956  &  19.50  &  0.708  &  21.10  &  0.747  &  22.45 \\
  1.155  &  16.90  &  0.953  &  19.60  &  0.700  &  21.15  &  0.753  &  22.50 \\
  1.144  &  17.00  &  0.951  &  19.70  &  0.693  &  21.20  &  0.759  &  22.55 \\
  1.131  &  17.10  &  0.948  &  19.80  &  0.687  &  21.25  &  0.765  &  22.60 \\
  1.120  &  17.20  &  0.946  &  19.90  &  0.683  &  21.30  &  0.771  &  22.65 \\
  1.109  &  17.30  &  0.944  &  20.00  &  0.679  &  21.35  &  0.777  &  22.70 \\
  1.099  &  17.40  &  0.943  &  20.05  &  0.677  &  21.40  &  0.783  &  22.75 \\
         &         &         &         &         &         &         &        \\
         &{\bf HB} &         &         &         &         &         &        \\
         &         &         &         &         &         &         &        \\
   V-I   &    V    &   V-I   &    V    &   V-I   &    V    &   V-I   &    V   \\ 
  0.050  &  19.22  &  0.175  &  18.31  &  0.300  &  18.09  &  0.425  &  17.95 \\
  0.075  &  18.82  &  0.200  &  18.25  &  0.325  &  18.06  &  0.450  &  17.93 \\
  0.100  &  18.61  &  0.225  &  18.21  &  0.350  &  18.03  &  0.475  &  17.91 \\
  0.125  &  18.48  &  0.250  &  18.16  &  0.375  &  18.00  &  0.500  &  17.88 \\
  0.150  &  18.38  &  0.275  &  18.13  &  0.400  &  17.98  &  0.525  &  17.86 \\
\end{tabular}
\end{minipage}
\end{table}

\section{The Age of Terzan 8}

A most efficient way to study 
the relative ages of clusters having similar metallicities is by
a detailed comparison of the main branches in the CMD.

In Figure 7, the ridge line of Terzan 8 is shifted by
 $\Delta V=-2.32$ and $\Delta (V-I)=-0.08$
to match the CMD of M68 
(Walker 1994) and by $\Delta V=-3.55$ and $\Delta (V-I)=-0.005$ 
to match the CMD of M55 (Ortolani \& Desidera, private communication).
Neither the Ter 8 ridge line nor the reference CMDs have been corrected
for reddening. However, we have taken E(B-V) values for M55 and M68 from the
compilation of Peterson (1993) to check if the adopted (V-I) shifts are 
consistent with the expected differences in color excess between Ter 8
and the reference clusters.
The color shift applied to Ter 8 relative to M68 is in excellent agreement
with the estimated difference in color excess between the two clusters
($\Delta E(V-I)_{M68-Ter8}=E(V-I)_{M68}-E(V-I)_{Ter 8} = - 0.091$). 
The same is found for Ter 8 relative to M55: the difference in color excess
($\Delta E(V-I)_{M55-Ter8}= - 0.013$) 
turns to be in good agreement with the applied shift
($\Delta (V-I)=-0.005$).

The Terzan 8 mean loci agree quite nicely with those of both clusters,
any subtle difference being attributable to unsignificant distortions 
of the shape of the ridge line. 

In order to study the age of Terzan 8 in more detail, we compute the
$\Delta V^{TO}_{HB}$ parameter, which is widely considered as one of the best
diagnostics to quantify age differences between GC pairs 
(see Buonanno, Corsi \& Fusi Pecci 1989; CDS, Stetson,
Vandenberg \& Bolte 1996).

From the mean ridge lines listed in Table 4, the TO point of Terzan 8 can 
be located at $V_{TO}=21.55\pm 0.10$; using the  $V_{HB}$ estimated
above, and propagating errors, we find
$\Delta V^{TO}_{HB}=3.60 \pm 0.11$. This value is compatible with that 
found for M55  ($\Delta V^{TO}_{HB}=3.5 \pm 0.1$ by Ortolani \& Desidera)
but slighlty greater than the figure determined by Walker (1994) for M68
($\Delta V^{TO}_{HB}=3.41 \pm 0.05$). Although the Terzan 8 and M68 
$\Delta V^{TO}_{HB}$ values 
agree to within the errors, the weak indication from Fig. 7
that M68 may be slightly younger than Terzan 8
is marginally supported.
Notwithstanding this small difference, which is of low significance given
the size of the errors, 
we can conclude that the derived value of
$\Delta V^{TO}_{HB}$ for Terzan 8 is in rather 
good agreement with those obtained for the typical old metal poor
GGCs. 
Thus, Terzan 8 is apparently coeval with the bulk of the 
galactic globulars of the same metallicity class, implying that the first
episodes of globular cluster formation
took place at the same time in the Milky Way and in the Sagittarius galaxy.

A similar conclusion has been reached from the study of the
oldest set of GC in the Large Magellanic Cloud (Testa et al. 1995,
Brocato et al. 1996). Moreover deep HST photometry of the globular clusters 
n. 1, 2, 3 and 5 in the Fornax dSph has been recently presented by Buonanno et
al. (1997). The ridge lines of the Fornax clusters were shown to match very
well the same CMDs of M68 and M55 we used here for comparisons with Ter 8,
so they too result coeval with ordinary {\it old} Galactic globulars. 
All these evidences suggests a scenario in 
which the formation of the oldest populations of clusters
started at nearly the same era (within $\sim 1~Gyr$) over a very wide volume 
around the Galaxy (see also Harris et al. 1997).

\begin{figure*}
\epsffile{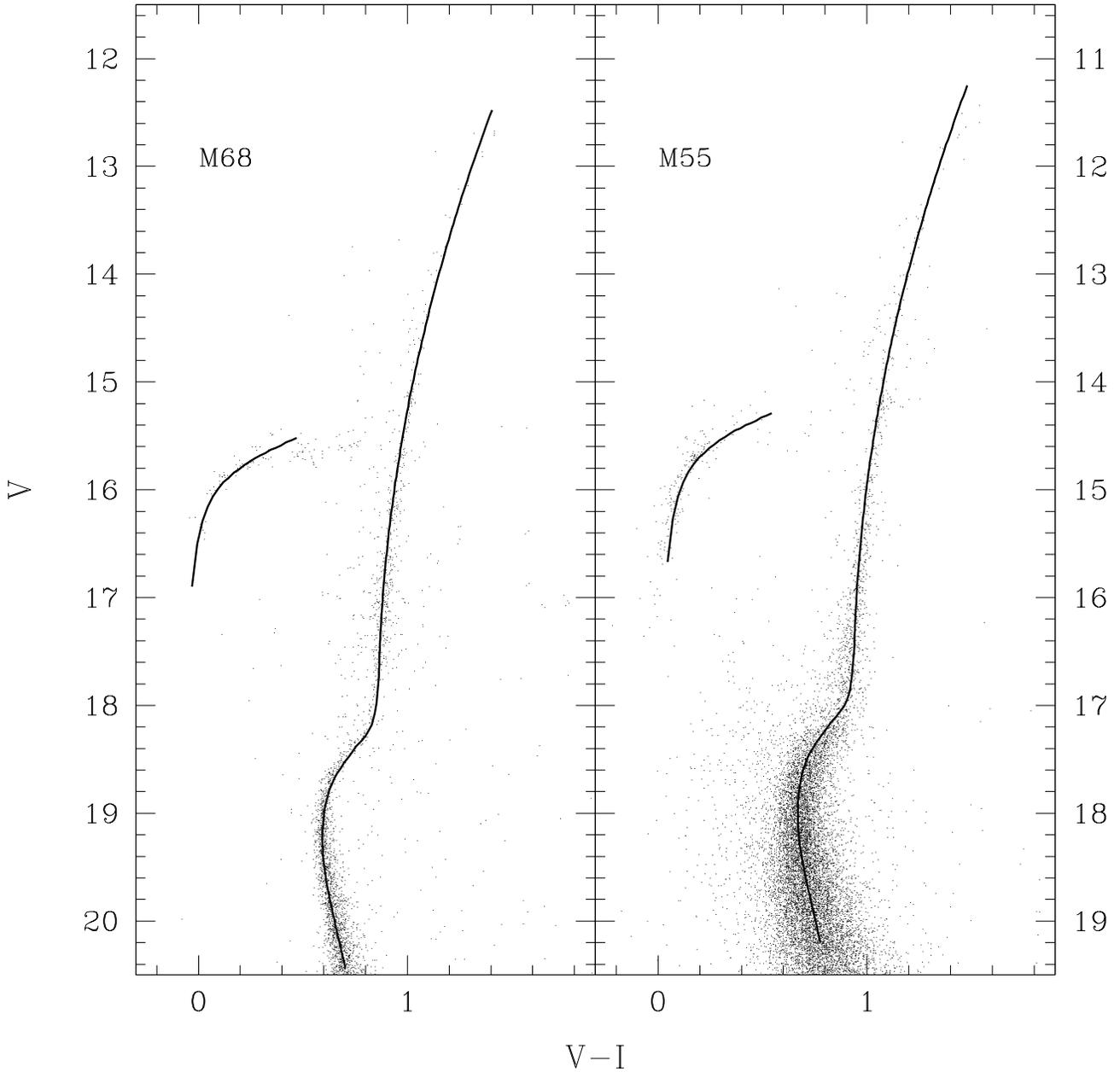}
 \caption{ The mean ridge line of Terzan 8 shifted 
($\Delta V=-2.32$ and $\Delta (V-I)=-0.08$) to match the CMD of M68 
(Walker 1994) {\it (panel a)}; and by 
$\Delta V=-3.55$ and $\Delta (V-I)=-0.005$
to match the M55 CMD by Ortolani \& Desidera }
\end{figure*}

\subsection{The Globular Cluster System of the Sagittarius Galaxy}

The Sgr and Fornax dSph galaxies share the property of having their
own globular cluster
systems. This is 
a very lucky opportunity to provide constraints on the Star Formation History
(SFH) of these galaxies, via age-dating of their clusters. While age
estimates of the Fornax GCs 
will be soon available from HST data (Buonanno et al. 1997)
reliable 
age determinations are indeed available
(from ground based observations)
 for all GCs belonging to the Sgr galaxy:
Arp 2, Ter 7 (Buonanno et
al 1995a,b), M54 (Marconi et al. 1997; Layden \& Sarajedini 1997), and 
Ter 8 (this paper).

So, the results from the relative age scale within the Sgr GC system 
provide the tantalizing possibility of studying its Age-Metallicity Relation
(even if it will appear somewhat noisy due to observational and 
theoretical uncertainties). 
Given the measured $[Fe/H]$ and $\Delta V^{TO}_{HB}$, we can derive the age
differences of the Sgr clusters with respect to Ter 8 which happens
to be the oldest and most metal-poor cluster of the group. This cluster
also provides a strong tie to the Galactic GC system time-scale, being
coeval with the oldest Galactic globulars.
In order to derive the age differences,
we use equations of the form $t_9 = f(\Delta V^{TO}_{HB}, [Fe/H])$ 
provided by CDS, applying their prescription for clusters with stubby red HB
morphologies to the case of Ter 7. 
For the sake of comparison
we calculate $\Delta age$ under two different assumption for the
slope $a$ of the relation between
the absolute magnitude of the HB and the cluster metallicity, 
i.e.:  $a=0.20$, corresponding to the second row of Table 1 of CDS and
$a=0.30$ corresponding to the last row of the same table (see 
Buonanno, Corsi, \& Fusi Pecci 1989, CDS, and
Vandenberg, Bolte \& Stetson 1996 for
discussion about the impact of $a$ on the Galactic age scales).  
It must be stressed that in the derivation of 
a {\it relative} age scale, only the {\it slope} of this relation is relevant.
 
Given the uncertainties in the observables and in the metallicity scale, the
adoption of a calibration based on different isochrones would not have a 
significant impact on the derived relative age scale. 

The results of our calculations are reported in Table
5, together with the metallicity and $\Delta V^{TO}_{HB}$ value
of each cluster, along with the associated
references. Two different entries are provided for the cluster Ter 7 reporting
the differences in its relative age due to the adoption of the
metallicity measurements obtained through spectroscopy of
giant stars or via photometric methods. 
This specific topic is widely discussed in
FBCF and Sarajedini \& Layden (1997) and we will not provide here any new 
insight.
Similar discrepancies are also found for M54 (Sarajedini \& Layden 1995, 1997)
and Arp 2 (FBCF), but they are only marginally significant. So we adopt, 
as a homogeneous set, the metallicity values derived via photometric 
methods (see Table 5 for references). It must be stressed that the
presented age-scales are also affected by this further uncertainty. 

In summary, any derived age scale is necessarily subject to 
{\em a)} errors in the observables, {\em b)} uncertainties in the
 slope of the $M_V(HB) ~vs.~ [Fe/H]$ relation, and {\em c)} 
uncertainties  associated with the choice of the metallicity 
({\em photometric vs. spectroscopic}), which is greater than the error of each 
single measure.
To take into account all these sources of uncertainty we assign a 
conservative error-bar of $\pm 2 Gyr$ to our relative age values.

\begin{figure}
\epsffile{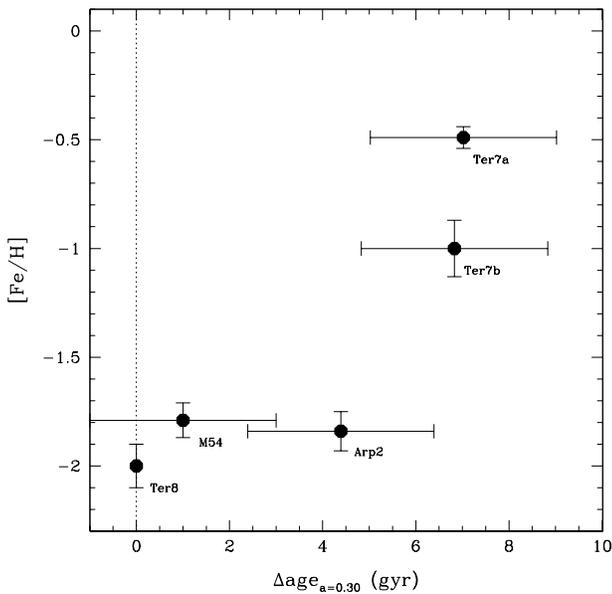}
 \caption{Age Metallicity diagram for the Sgr globulars. The metallicity
error bars are from the source indicated in Table 5. 
The errors bars in age have been assumed largely conservative
($\pm 2$ Gyr) in order to account for all the effect quoted in section 5.1.
The points corresponding to both the Ter 7 entries of Table 5 are displayed
(see sec 5.1).}
\end{figure}

With this assumpion, the two age-scales do not show significant differences.
In the following
discussion we will thus adopt the age-scale associated with the higher slope of
the  $M_V(HB) ~vs.~ [Fe/H]$ relation (i.e. $a=0.30$; column \#5
 of Tab. 4) as a conservative choice, being the one which minimizes the
 age differences between clusters.

\begin{table*}
 \centering
 \begin{minipage}{140mm}
  \caption{Relative age scales for the Sgr GC System}
  \begin{tabular}{@{}lccccc@{}}
Cluster & [Fe/H]  & $\Delta V^{TO}_{HB}$ & $\Delta age_{a=0.20}$ & 
$\Delta age_{a=0.30}$& Ref.\\
Ter 8 & $ -2.00 \pm 0.10$ &  3.60  &  0.00  &  0.00 & present paper \\
Ter 7 & $ -0.49 \pm 0.05$ &  3.20  &  9.75  &  7.02 & B95a + S95\\
Ter 7 & $ -1.00 \pm 0.13$ &  3.20  &  8.86  &  6.83 & B95a\\
Arp 2 & $ -1.84 \pm 0.09$ &  $3.29^{a}$  &  5.53  &  4.39 & B95b and SL97\\
M 54  & $ -1.79 \pm 0.08$ & $3.55^b$  &  1.50  &  1.00 & SL95 and LS98\\
\end{tabular}

{\footnotesize B95a = Buonanno et al. 1995a; B95b = Buonanno et al. 1995b;
M97 = Marconi et al. 1997;
S95 = Suntzeff et al. 1996; SL95 = Sarajedini \& Layden. 1995; 
SL97 = Sarajedini \& Layden 1997; 
LS98 = Layden \& Sarajedini 1998;

$^a$ CDS erroneously interpreted the
$V_{HB}$
value provided by Buonanno et al. 1995b for this cluster as the V magnitude
of the Zero Age Horizontal Branch and, consequently, applied an unnecessary 
correction to obtain $V_{RR}$ for this cluster, since $V_{HB}$ was instead 
estimated in a way very similar to the one applied in the present paper
to Ter 8 (Buonanno et al., private communication). 

$^b$ This is preferred over the M97 value because the photometry of 
LS98 (see also Layden \& Sarajedini 1997) extends deeper than that
of M97. However, within the errors, the $\Delta V^{TO}_{HB}$ values
of M97 and LS98 are in reasonable agreement.}
\end{minipage}
\end{table*}

\subsection{The Age - Metallicity Relation}

Figure 8 displays the {\it observed} Age - Metallicity Relation (AMR) 
for the Sgr GC System, as derived with the prescription described above. 

It is worth stressing that significant features are present in the diagram 
despite the large uncertainties involved and {\em independent} of the
assumptions inherent in the adopted $t_9 = f(\Delta V^{TO}_{HB}, [Fe/H])$ relation:
\begin{enumerate}

\item{} Arp 2 and Ter 7 are certainly younger than Ter 8, i.e. overall 
globular cluster formation lasted for $> 4 ~Gyr$ in the \sgr.

\item{} Ter 7 is at least three times more metal rich than any other Sgr
cluster, i.e. a great deal of chemical enrichment occurred in the lapse of 
time during which all the Sgr globulars (and probably also the bulk of its
field stars; see below) were formed.

\end{enumerate} 

Besides this, coupling the information contained in Figure 8 with the results
of many recent studies of the Sgr galaxy and its clusters (see Ibata et al. 1997
and Layden \& Sarajedini 1997 for references), we note that: 

\begin{itemize}

\item{}The presence of an old stellar population (i.e. of age comparable to 
Ter 8) in the Sagittarius galaxy is confirmed by the identification of RR Lyrae
variables (Mateo et al. 1995; Alard 1996; Alcock et al. 1996) and Planetary 
Nebulae (Zijlstra \& Walsh 1996), while the dominant population looks very 
similar to Ter 7, both in age and metallicity (Marconi et al. 1997). Any
possible younger population, whose existence has been suggested by Mateo et al.
(1995), Layden \& Sarajedini (1997) and Whitelock et al. (1996), is 
constrained by the present data to contribute only a small fraction of Sgr 
stars. So, it seems reasonable to assume that the AMR presented in 
Figure 7 has to track the main features of the AMR of the whole galaxy.

\item{} From Figure 8 and from the above dicussion it is apparent that
the \sgr formed relatively metal poor stars and clusters for a significant
fraction of its evolutionary history, but the main star formation 
event (i.e. the
one associated with Ter 7) occured at a time when the interstellar medium had
been greatly enriched in heavy elements. The process of enrichment could have 
been completed in $\sim 1-2 ~Gyr$, as judged from the age difference between
Arp 2 and Ter 7. 

\item{} Massive star formation in the \sgr ceased several gigayears ago. Even if
sporadic episodes occured later than the epoch during which Ter 7 formed, no
hint has been found for the existence of populations younger than $\sim 4 ~gyr$,
and the system is presently found to be totally gas-depleted (Koribalski, 
Johnston \& Optrupceck 1995). Coupled
with the fact that the orbital path of Sgr brings it very near the 
Galactic bulge/disk (Ibata et al. 1997), this evidence strongly suggests that 
the star formation history of this galaxy has been largely influenced 
by some episode of gas-stripping due to interactions with the Milky Way.
It seems also likely that the major mass contribution
injected in the Galactic Halo and/or Disk by the Sgr system was in the
form of enriched gas (presumably at the metallicity of Ter 7), not
stars (see also Gilmore 1996).

\item{} A potentially powerful test to allow better time resolution at older
ages would be to measure the $[\alpha/Fe]$ ratio both in the GCs and the field
stars of Sgr. It is reasonable to assume that, in the presence of previous 
star formation,
clusters formed after the enrichment of the medium by type Ia supernovae
(a time scale generally believed to be $\sim 1 ~gyr$; Wheeler, Sneden \&
Truran 1989)
should display a lower $[\alpha/Fe]$ ratio with respect to typical Galactic
Halo GCs and stars (Wheeler, Sneden \& Truran 1989; Cayrel 1996), while 
``ordinary'' {\it $\alpha$-enhancement} would imply formation within 
$\sim 1 ~Gyr$ from the onset of star formation in the system. The possibility
that the Sgr globulars can present differences in abundance patterns has been
already suggested and discussed by FBCF and Sarajedini \& Layden (1997).  

\item{} If Sgr has to be considered a typical SZ ``fragment'',
which is being accreted and is in the process of building up the Galactic 
Halo, the above 
considerations show that such systems 
may leave signatures in the
ordinary Galactic Halo in the form of 
anomalies in the age and metallicity distributions of Halo globulars and field
stars. 

\end{itemize}

\section*{Acknowledgments}
This research has made use of NASA's Astrophysics Data System Abstract Service. 
We thank S. Ortolani and S. Desidera for kindly sending us the 
data on M55 before publication.
Paolo Montegriffo was supported 
by a 1996-grant of the {\it Fondazione del Monte, Rolo Banca 1473}.
Donald Martins was supported by a research and travel grant from the University
of Alaska, Anchorage, and would like to acknowledge the friendly
hospitality of the National Optical Astronomy Observatories during his
visit there. Ata Sarajedini was supported by the National Aeronautics and 
Space Administration (NASA) grant number HF-01077.01-94A from
the Space Telescope Science
Institute, which is operated by the Association of Universities for
Research in Astronomy, Inc., under NASA contract NAS5-26555.
The finacial support of the {\it Ministero delle Universit\`a e della
Ricerca Scientifica e Tecnologica} (MURST) and of the {\it Agenzia
Spaziale Italiana} (ASI) is kindly acknowledged.

We are also indebted to an anonymous referee for the many useful comments.

\end{document}